\documentclass[aps,%showpacs,
twocolumn,toolkits,nofootinbib]{revtex4}
\usepackage{graphicx,epsf,amssymb,amsbsy,amsfonts,amssymb,amsmath,mathtools}
\usepackage[hidelinks]{hyperref}
\usepackage[usenames]{color}

\newcommand{\eq}{\begin{equation}}
\newcommand{\eqe}{\end{equation}}

\newcommand{\eqa}{\begin{eqnarray}}
\newcommand{\eqae}{\end{eqnarray}}

\newcommand\eea{\end{eqnarray}}
\newcommand\bea{\begin{eqnarray}}

\def\d{\partial}
\def\<{\langle}
\def\>{\rangle}
\def\+{\dagger}

\def\I2{\mathbb{I}_2}
\def\sech{{\rm sech}}

\begin{document}

\title{Short-distance Schwinger-mechanism and chiral symmetry}

\author{Jon Brog\aa rd}

\author{David A. McGady}
\email{mcgady@nbi.ku.dk}

\affiliation{Niels Bohr International Academy \\
17 Blegdamsvej K\o benhavn 2100, Denmark}

\begin{abstract}
In this paper, we study Schwinger pair production of charged massless particles in constant electric fields of finite-extent. Exploiting a map from the Dirac and Klein-Gordon equation to the harmonic oscillator, we find exact pair production rates for massless fermions and scalars. Pair production rates depend only on the ratio between the capacitor plate separation, $\ell$, and the length-scale of the force-field, $\ell_F$. Chirality ensures that fermion production smoothly vanishes with $\ell/\ell_F$. Scalar pair production though diverges exponentially quickly in this limit. The same limit of the smooth tanh-potential does not diverge; divergences seem tied to singularities in current and charge densities. 
\end{abstract}

\maketitle

\section{Schwinger pair creation}

Effectively, a system at zero-temperature can be said to be in its vacuum-state if it contains no particles that propagate over distances comparable to its size. However, this definition is a size-dependent statement: it is insensitive to fluctuations of matter-fields that propagate over distances and times that are much smaller than the system size. Particles that propagate only short distances correspond to off-shell fluctuations of matter fields. When deviations in energy and momenta $\Delta E \!\sim\! \Delta p$ carried by these virtual states are small compared to the characteristic distance-scale over which they propagate $\Delta t \!\sim\!\Delta x$, these fluctuations are not directly unobservable: Fundamentally, $\Delta x \Delta p \lesssim 1$, and the vacuum remains the vacuum. 

However, as Schwinger famously observed~\cite{01-Schwinger} powerful electric fields can qualitatively change this picture: If an electric field can do work comparable to the rest-mass of a virtual particle-antiparticle pair, $2m \lesssim q E d$, over a distance that is smaller than the pair's Compton-wavelength, $d \lesssim 1/m$, then the pair can be driven onto its mass-shell and propagate over long distances and times. Thus, if the strength of the electric field $E$ satisfies
\begin{align}
q E \gtrsim 
m^2~,
\end{align}
then electrically charged particle-antiparticle pairs will be spontaneously created from what would otherwise be the vacuum-state of the charged-matter field. This effect is referred to as the Schwinger mechanism, though it has earlier precedents in the literature~\cite{02-EulerHeisenberg}.

More precisely, in~\cite{01-Schwinger}, Schwinger calculated the one-loop effective action for a charged fermion propagating in an external electromagnetic field in $(3+1)$-dimensions. When the field is a homogeneous electric field, Schwinger found that the effective Lagrangian acquires a non-zero imaginary part given by,
\begin{align}
2~ {\rm Im}~{\cal L}_{1-{\rm loop}} = \frac{(q E)^2}{4 \pi^2} \sum_{n = 1}^{\infty} \frac{1}{n^2} e^{-n\big(\tfrac{\pi m^2}{q E}\big)} ~.
\end{align}
Because of this, the vacuum-to-vacuum amplitude in a volume of spacetime $VT$ must exponentially decays as ${\rm Exp}[-V T {\rm Im}{\cal L}]$. Vacuum decay, here, directly corresponds to the production of electron-positron pairs from the electron-positron vacuum-state. This mechanism is fundamentally related to many important aspects of quantum field theory, and has recently been proposed to be directly observable in the condensed-matter setting of graphene~\cite{03-SchwingerGraphene}. 

Theoretically, Schwinger pair creation from electric fields is similar in many ways to Hawking pair-creation via gravitational field gradients at the event horizons of black-holes~\cite{04-Hawking}. For this reason, Schwinger pair creation and the ensuing back-reaction against the applied electric field is sometimes regarded as a simple analog to Hawking evaporation. See for example Ref.~\cite{05-HS-merge1} and Refs.~\cite{06-HS-merge2, 07-HS-merge3, 08-HS-merge4}. It has further been used to study the physics of the QCD-string, a phenomenological model of low-energy quark-antiquark interactions where the interaction governed by the confining and linearly increasing potential $V(x) = F x$. This potential is directly analogous to the linearly-increasing scalar potential between two capacitor plates. In this context, Schwinger pair-creation constitutes parton-parton pairs breaking the QCD flux-tube, creating parton-showers with low energy and transverse momenta~\cite{09-QCDstring1, 10-QCDstring2}. 

Schwinger pair creation can be understood as a tunneling process. In the gauge where an electric field is given by the gradient of a scalar potential, $\d_x A_0(x)$, the potential shifts the Dirac-Fermi sea-level on one side of the electric field relative to the other. In this gauge, the filled negative-energy states below the Dirac-Fermi sea on one side of the potential become degenerate to the empty positive-energy states on the other side of the potential. Thus, pair production corresponds to a filled negative-energy state tunneling to become a filled positive-energy states on one side while leaving behind an empty negative-energy state on the other.\footnote{Alternatively, in the gauge where $\vec{E}(t,\vec{x}) = \d_t \vec{A}(t,\vec{x})$, Schwinger pair creation corresponds to Landau-Zener tunneling~\cite{11-LandauZener1, 12-LandauZener2} between occupied negative-energy states and empty positive-energy states. This statement connects the Landau-Zener and Klein-tunneling literature in graphene~\cite{13-KleinGraphene1, 14-KleinGraphene3, 15-KleinGraphene3} to the proposed experimental test of the Schwinger mechanism in graphene in Ref.~\cite{03-SchwingerGraphene} and Refs.~\cite{16-SchwingerRevisit, 17-SchwingerLZ}.}

This picture is helpful quantitatively and qualitatively, Qualitatively, one can compute the approximate amplitude that a filled negative energy state with mass $m$ and momentum transverse to the field direction $\vec{p}_T$. Semiclassical/WKB methods yield~\cite{16-SchwingerRevisit, 18-Nikishov}:
\begin{align}
\frac{d N}{dt ~ dE ~ d^{d-2}V_T ~ d^{d-2}p_T} \simeq |T|^2_{\rm WKB} = e^{-\pi \tfrac{m^2 + p_T^2}{qE}}~.
\end{align}
Quantitatively, there are several prominent examples where the tunneling amplitude can be computed exactly. In this paper, we will be chiefly interested in the exact pair-production rate in electric fields whose intensity varies as $F \sech^2(x/w)$~\cite{18-Nikishov,19-Sauter1,20-Sauter2} and in constant electric fields of finite spatial extent~\cite{21-WangWong}. It is important to mention Refs.~\cite{22-SchwingerQFT1, 23-SchwingerQFT2}, which discuss how Schwinger pair production framed in terms of tunneling in relativistic wave equations can be formulated in a fully field-theoretic language.

The purpose of this paper is to study deviations from the semiclassical Schwinger rate, due to short-distance effects. All of our calculations are done for the two exactly solvable potentials described above. We are particularly interested in how pair-production in a constant electric field of finite spatial extent depends that extent. In addition to being exactly solvable~\cite{16-SchwingerRevisit, 18-Nikishov, 21-WangWong}, this field configuration is a good approximation to electric fields generated by parallel-plate capacitors. Each fact makes pair creation in constant electric fields highly relevant for the experimental test of the Schwinger mechanism in graphene~\cite{03-SchwingerGraphene}.

Motivated by graphene, where the charged electron quasiparticles behave as if they are massless and obey the (2+1)-dimensional Dirac equation, we study Schwinger pair creation of \emph{massless} charged particles in these exactly solved potentials. Creating a pair in a given mode can be mapped to tunneling in (1+1)-dimensions. We further focus on potentials with exactly solved transmission coefficients. When the particles are massless, the total pair creation rate can only depend on the dimensionless parameter $\ell/\ell_F$, where $\ell$ is the width of the field region and $\ell_F$ is the length-scale associated with the electric field, $\ell_F:= (qE)^{-1/2}$. 

Intriguingly, even though the exact pair production rates in a constant electric field of finite width have been known for nearly thirty years~\cite{21-WangWong}, it seems that the regime $\ell/\ell_F \to 0$ has not been well-explored. However, the the recent proposed experimental test of the Schwinger mechanism in graphene in Ref.~\cite{03-SchwingerGraphene} suggests that this region of parameter-space may be experimentally relevant. Our main result is that while the total fermion-production rate smoothly vanishes as $\ell/\ell_F \to 0$, the rate for scalar production \emph{diverges} faster than any power-law in $\ell_F/\ell$ when $\ell/\ell_F \to 0$ (Fig.~\ref{fig6}). Divergent scalar production in this limit seems to have escaped prior notice in the literature. In contrast, massless fermionic production is naturally protected by chiral symmetry and Fermi-Dirac statistics: it saturates at unity.

The structure of the paper is as follows. In sections~\ref{secCap},~\ref{secDivergence} and~\ref{secSmoothing}, we study linear and $\tanh$-potentials as $\ell/\ell_F  \to 0$ while holding $F = \ell_F^{-2}$-fixed. We conclude in section~\ref{secEnd}.

\section{Capacitors and Integrability}\label{secCap}

In this section, we show why tunneling amplitudes are exactly solvable for constant electric fields. Simply, constant electric fields are given by the first derivative of linearly increasing vector-potentials, $\d A =$ constant. Because relativistic wave equations are quadratic in the covariant momenta, $(p - A)^2$, the normal quadratic terms in the naked spatial variables now come paired with quadratically increasing functions of the coordinates. Finally, in the one-dimensional transmission problems of interest the dynamics distill down to solutions of a one-dimensional differential equations of the schematic form,
\begin{align}
\big[ (E - Fx)^2 - p_x^2 - \Lambda \big] \psi(x) = 0~,
\end{align}
where $F$ is the strength of the electric field and $\Lambda = m^2 + p_T^2$ is a constant that depends on $m^2$, the mass of the particle, and $p_T^2,$ the momentum of the scattering state perpendicular to the homogeneous field. Crucially, this differential equation is an analytic continuation of the Schrodinger equation for the harmonic oscillator: The differential operator is changed from $(x^2-\d_x \d_x)$ to $-(x^2+\d_x \d_x)$, and now has continuous eigenvalues $n + 1/2 \to m^2 + p_T^2$

Thus, there is a hidden algebra of creation and annihilation operators that dictate the local solutions to the Klein-Gordon and Dirac equations.\footnote{The harmonic nature of solutions to the wave equations should be visible by analytic from the Landau Level problem, as well. In the presence of a constant $B$-field, the Landau-level solutions are deeply related to solutions to the harmonic oscillator. Relativistically, we can analytically continue $F_{\mu \nu}^2 \propto E^2 - B^2$ from $-B^2$ to $+E^2$ by sending $B \to i B$ and then defining $E = i B$.} This algebra is the reason why these relativistic wave-equations are exactly solvable in terms of special functions, and gives the results from our analysis added robustness. It is important to see how the Klein-Gordon and Dirac equations manifest this harmonic structure. We do this in three steps.

First, we work explicitly in (1+1)-dimensions. As discussed in~\cite{16-SchwingerRevisit}, one can derive the full pair creation rates and vacuum decay rates in $d$-dimensions by treating the field direction as privileged and simply considering the transverse directions by studying the tunneling problem as mode-by-transverse-mode tunneling effect in (1+1)-dimensions. The essential dynamics are completely encoded in the (1+1)-dimensional problem. 

Second, we work with an electric field associated with parallel capacitor: it is zero for $|x| > \ell/2$ and constant for $|x| < \ell/2$. Third, we write the field as the gradient of the scalar potential
\begin{align}
&A_0(x)=   
\begin{cases} 
F\ell/2		& {\rm for} ~ x \leq -\ell/2  \\ 
-Fx		& {\rm for} ~ \leq x \leq \ell/2 \\ 
-F\ell/2		& {\rm for} ~ \ell/2 \leq x 
\end{cases} \nonumber \\
\implies 
&F(x)=   
\begin{cases} 
0 	& {\rm for} ~ x\leq -\ell/2 \\ 
F 	& {\rm for} ~ -\ell/2 \leq x \leq \ell/2 \\ 
0 	& {\rm for} ~\ell/2 \leq x 
\end{cases}~. \label{eqForceDef}
\end{align}
There are three regions: region-I has $x < -\tfrac{\ell}{2}$, region-II has $|x| \leq \tfrac{\ell}{2}$ and region-III has $\tfrac{\ell}{2} < x$. 

In this gauge the relativistic wave equations are then invariant under time-translations. Therefore energy is conserved for scattering states, which take the form
\begin{align}
\phi(x,t) = e^{-i E t} \Phi(x)~.
\end{align}
Using this, we now write the Klein-Gordon and Dirac equations, 
\begin{align}
\big[ \big(p - A(x,t)\big)^2 - m^2 \big] \phi(x,t) &= 0 ~,~~ {\rm and} \nonumber\\  
\big[ \big(\slash \!\!\!\! p - \slash \!\!\!\! A(x,t) \big) - m \big] \psi(x,t) &= 0~,
\end{align}
so that they manifestly resemble the harmonic oscillator. It is helpful to use the variables
\begin{align}
\ell_F := 1/\sqrt{F} \quad {\rm and} \quad 
\mu := -\ell_F m \quad , \quad 
\chi:= \ell_F E + x/\ell_F~.
\end{align}
With these variables, and in this gauge, the Klein-Gordon and Dirac operators acting on their fixed-$E$ eigenfunctions reduce to,
\begin{align}
\big(\d_{\chi}^2 + \chi^2 - \mu^2\big) f(\chi) &=  0~,
\quad {\rm and} \nonumber\\
\left( \begin{matrix} 
\mu & \chi - i \d_{\chi}   \\ 
\chi + i \d_{\chi} & \mu
\end{matrix} \right) 
\left( \begin{matrix} 
\psi_{+}(\chi) \\ \psi_{ -}(\chi) 
\end{matrix} \right) 
&= 0~, \label{eqODE1}
\end{align}
where we have used the chiral basis $\gamma_0 = i \sigma_2$ and $\gamma_1 = \sigma_1$. We now recall the precise form of the classic creation and annihilation operators for the harmonic oscillator $(p_y^2 + y^2 +1)/2$:
\begin{align}
a_y^{\pm} &= \frac{1}{\sqrt{2}}(y \pm i p_y) = \frac{1}{\sqrt{2}}(y \mp \d_y ) \\ \nonumber
\implies [a_y^{-},a_y^{+}] &= 1 \quad {\rm and} \quad H = \frac{a^- a^+ + a^+ a^-}{2}~.
\end{align}
Identifying $\chi := y\sqrt{i}$ implies $p_{\chi} = p_y/\sqrt{i}$, and allows us to write the differential operators for the Klein-Gordon and Dirac equations in terms of the complexified analogs of $a^{\pm}_y$,
\begin{align}
a_{\chi}^{\pm} = \frac{1}{\sqrt{2}}(\chi \pm p_{\chi}) = \frac{1}{\sqrt{2}} (\chi \pm i \d_{\chi}) 
\implies [a_{\chi}^{-},a_{\chi}^{+}] = i~.
\end{align}
With this, the Klein-Gordon (KG) and Dirac equations respectively become
\begin{align}
{\rm KG}:
~\big( a_{\chi}^{+} a_{\chi}^{-} + a_{\chi}^{-} a_{\chi}^{+}-(i + \mu^2) \big) f(\chi) &= 0~, 
~~ {\rm and} \nonumber\\
{\rm Dirac}:
~\left( \begin{matrix} 
\mu & a_{\chi}^{-}   \\ 
a_{\chi}^{+} & \mu
\end{matrix} \right) 
\left( \begin{matrix} 
\psi_{+}(\chi) \\ \psi_{ -}(\chi) 
\end{matrix} \right) &= 0~.
\end{align}
In the chiral limit $\mu \to 0$; the Dirac equation for massless fermions in (1+1)-dimensions is entirely written in terms of the creation and annihilation operators. Thus, spinor-components are simply eigenfunctions of the harmonic oscillator. This will be important below.

\section{Small distances and masslessness}\label{secDivergence}

In the previous section, we explained why the Klein-Gordon and Dirac equations are ``exactly'' solvable in terms of (analytic continuations of) elementary known special functions. Having established the origin of the exact solvability for the potential of interest, we now use it to derive exact results for Schwinger pair creation from this field. To be clear, these results are exact only in the limit where we can safely neglect back-reaction. Back-reaction is a crucially important feature of Schwinger pair creation that imposes limits on the current experiment proposed in graphene~\cite{03-SchwingerGraphene}, and is of great theoretical interest~\cite{05-HS-merge1,06-HS-merge2, 07-HS-merge3, 08-HS-merge4, 24-Polyakov1, 25-Polyakov2}.

Our analysis in this section has four aspects. First, we set-up the exact scattering problem for scalars and for fermions subjected to an external electric field given by parallel capacitor plates separated by a distance $\ell$ (and unbounded in the transverse directions). Second, we exactly solve for fermion and scalar pair production in this field configuration. Crucially, for finite masses and widths we see excesses and deficits relative to the classic Schwinger rate.

Third, we show that as we approach the massless limit, fermion pair production never exceeds the Schwinger result, and is always finite. However, as the scalar mass goes to zero, the surplus above the Schwinger rate remains. Strictly at $m = 0$, there are only two length-scales: the separation between the capacitor plates, $\ell$, and the length-scale of the force-field $\ell_F$. Here, the Schwinger rate depends only on the dimensionless ratio $\ell/\ell_F$. Importantly, we see that when $\ell/\ell_F \to 0$ the scalar production rate diverges faster than any power-law.

Fourth, we show why the fermion pair production remains finite in the chiral limit. There are two ways to see this physically. First, as $m \to 0$ scattering states have well-defined chirality. This suppresses reflection/back-scattering; interactions with the electric field conserve helicity. Helicity conservation helps to trivialize and unitarize fermion scattering and production. Second, the Schwinger rate saturates at unity when $m^2 = 0$. Fermi-Dirac statistics apply, and occupation numbers cannot exceed unity, implying that the Schwinger rate cannot be exceeded. (The algebra of creation and annihilation operators dominate in the chiral limit; this also explains saturation.) Neither feature, however, governs scalars. It is thus unsurprising that scalar pair creation rates exceed the Schwinger rate, and even diverge.

\subsection{Scalar production from capacitors}~\label{secScalarProd}

In this section, we derive the precise form of the transmission coefficients for charged scalars in a constant electric field of finite width. We work in the gauge $A_{\mu}(x_{\nu}) = (A_0(x),0)$ and focus on the $A_0(x)$ defined in Eq.~\eqref{eqForceDef}. In this gauge, energy is well-defined as both the constraints and the equations of motion are time-translation invariant. However, the scalar potential raises the effective Fermi-Dirac sea-level on one side of the field region relative to the other. This makes filled negative energy states on one side degenerate with empty positive energy states on the other, when $|E| \leq \ell/2\ell_F^2 -m$. Pair production thus occurs when filled negative energy states in this mixing region tunnel through the field and populate the empty positive energy states on the other.

This transmission coefficient $T$ depends on the energy of the state in this band, the width of the electric field region, and the strength of the electric field. Thus $T$ is more properly written $T(E,\ell,\ell_F,m)$. To obtain it, we must match the local solutions and their first derivatives on the boundary between the plane-wave solutions on either side of the capacitor and the non-trivial solutions to the Klein-Gordon equation~\eqref{eqODE1} in the field-region. 

Outside the capacitor, namely in region I with $x<-\tfrac{\ell}{2}$ and in region III with $\tfrac{\ell}{2} < x$, the solutions to the Klein-Gordon equation are  plane-waves:
\begin{align}
\phi_{\rm I}(x,t) &= e^{-i E t} \left[ \frac{e^{- i x |k_-|x}}{\sqrt{|E_{-}|}} + R \frac{e^{+ i x |k_-|x}}{\sqrt{|E_{-}|}} \right] \\
\phi_{\rm III}(x,t) &= e^{-i E t} \left[ T \frac{e^{+i x |k_+|x}}{\sqrt{|E_{+}|}} \right]~,
\end{align}
where the momenta $k_{\pm}$ are given in terms of the locally shifted energies $E_{\pm}$:
\begin{align}
&k_{\pm} := \sqrt{E_{\pm}^2-m^2} ~,~ \\
&E_{\pm}:= E- A_0\big(x \geq \pm|\tfrac{\ell}{2}|\big) = E \pm \tfrac{\ell}{2 \ell_F^2} 
\end{align}
on either side of the electric field defined in Eq.~\eqref{eqForceDef}.

Inside the capacitor, in region II  with $|x| \leq \tfrac{\ell}{2}$, the solutions to the Klein-Gordon equation~\eqref{eqODE1} are given by Hermite polynomials,
\begin{align}
\!\!\!\! \!\!\!\! 
&\phi_{\rm II}(x,t) = e^{-i E t} \left[ \sum_{\lambda = \pm} A_{\lambda} \phi_{\lambda}(x|E,\ell_F,m) \right] \label{eqSmid1a} \\
&\phi_{\lambda}(x|E,\ell_F,m) := s_{\lambda}(\chi,\mu)~  ,
\!\!\!\! 
\label{eqSmid1}  \\
&s_{\pm}(\chi,\mu) = 
\sqrt{2}^{\tfrac{1}{2}\pm \tfrac{i}{2} \mu^2 } 
{\rm Exp}\left[ -\tfrac{1}{2} (e^{\pm i \pi/4}\chi)^2\right] \nonumber\\
&\qquad \qquad \qquad \qquad  {\rm H}\!\left[-(\tfrac{1}{2}\pm \tfrac{i}{2} \mu^2), e^{\pm i \pi/4} \chi \right]~,\!\! \label{eqScalarSol}
\end{align}
where again $\mu = m \ell_F$, $\chi = (E \ell_F + x/\ell_F)$, $A_{\pm}$ are constants fixed by the matching conditions at $x = \pm \ell/2$, and ${\rm H} [a,b]$ is an index-$a$ Hermite polynomial in variable-$b$. 

Replacing $i + \mu^2 \to 2n+1$ and $e^{i \pi/4} \chi \to x$ changes $s_{-}(\chi,\mu)$ into $\< x|n\> = {\rm H} [n,x]$, the $n^{th}$ energy eigenstate of the harmonic oscillator. This transformation maps $s_{+}(\chi,\mu)$ into the other local solution to the harmonic oscillator's Schrodinger equation, ${\rm H} [-(n+1),i x] = \< x|-(1+n)\>$, which grows exponentially with $x$. Physically, eigenfunctions for the harmonic oscillator must have a finite norm when integrated over the infinite range of the potential. Thus, $\< x| -(n+1)\>$ is not a valid solution to the harmonic oscillator. 

However, in our scattering experiment, the (inverted) quadratic potential has finite range. For this reason, both of the local solutions $s_{\pm}$ have finite norm, and so both of the $A_{\lambda}$ in Eq.~\eqref{eqSmid1a} are nonzero. We must keep both solutions in our description of this scattering process. Equipped with the explicit solutions in the three regions, $\phi_{\rm I}, \phi_{\rm II}$ and $\phi_{\rm III}$, we find the \emph{four} undetermined parameters $T$, $R$ and $A_{\pm}$ by solving,
\begin{align}
\begin{cases}
\phi_{\rm I }(x =  -\tfrac{\ell}{2})~		= \phi_{\rm II }(x =  -\tfrac{\ell}{2}) \\ 
\d_x \phi_{\rm I }(x = -\tfrac{\ell}{2})~ 	= \d_x \phi_{\rm II }(x =  -\tfrac{\ell}{2}) \\
\phi_{\rm II}(x = +\tfrac{\ell}{2}) 		= \phi_{\rm III}(x = +\tfrac{\ell}{2})  \\
\d_x \phi_{\rm II}(x = +\tfrac{\ell}{2}) 	= \d_x \phi_{\rm III}(x = +\tfrac{\ell}{2}) 
\end{cases}~.
\end{align}
This exactly reproduces the transmission coefficient $T$ found in~\cite{21-WangWong}:
\begin{align}
T(E)_{\rm scalar} = \frac{4 i k_{-} e^{\tfrac{i}{2} \ell (k_{\rm I} - k_{\rm III})} \sqrt{\tfrac{|E_{+}|}{|E_{+}|}}}{ {\cal D}\phi_{+}(+\tfrac{\ell}{2}) {\cal D}\phi_{-}(-\tfrac{\ell}{2}) - {\cal D}\phi_{-}(+\tfrac{\ell}{2}) {\cal D}\phi_{+}(-\tfrac{\ell}{2})}~, \label{eqScalarT}
\end{align}
where one of the factors of $2$ in the numerator comes from the Wronskian of $\phi_+$ and $\phi_-$, and ${\cal D}\phi_{\lambda}(x)$ is like a covariant derivative of $\phi_{\lambda}(y|E,\ell_F,m)$ evaluated at the position $x = y$:
\begin{align}
{\cal D}\phi_{\lambda}(x) := \d_y \phi_{\lambda}(y|E,\ell_f,m) \big|_{y = x} - i k_{\lambda} \phi_{\lambda}(x|E,\ell_F,m)~. \label{eqDerivative}
\end{align}
As shown in~\cite{21-WangWong}, when the width between the capacitor plates is large, the exact pair creation rates that come from these tunneling amplitudes match the Schwinger rate. Explicitly,
\begin{align}
\lim_{\ell/\ell_F \to \infty} |T(E,\ell,\ell_F,m)|^2 = e^{-\pi (m \ell_F)^2} = e^{-\pi \tfrac{m^2}{F}}~.
\end{align}
However, the transmission coefficient in Eq.~\eqref{eqScalarT} provides exact corrections to the semiclassical Schwinger result which come from finite-distance corrections. 

In particular, we trust the Schwinger result when semiclassical analysis is valid. This happens when the separation between the capacitor plates is large compared to the characteristic length-scale of the electric field and large compared to the Compton wavelength of charge-particles at rest. Thus, we should only expect the Schwinger rate to hold when $\ell \gg \ell_F$ and $\ell \gg 1/m, 1/|E_{\pm}|$. However as Ref.~\cite{21-WangWong} emphasizes, the tunneling rates in Eqs.~\eqref{eqScalarSol} and~\eqref{eqScalarT} are exact for all values of any dimensionless combinations of $\ell$, $\ell_F$, $m$, and $E$.

\subsection{Fermions, helicity, and null states}~\label{secFermionProd}

We now turn our attention to determining the charged fermion transmission amplitudes and pair creation processes. The setup is largely the same as for charged scalar transmission amplitudes, with the exception that the transmission amplitude is obtained by matching the two spinor components at each of the two boundaries $x = \pm \ell/2$. Again, the range of energies where pair production can occur is given by $-(\ell/2 \ell_F^2-m) \leq E \leq (\ell/2 \ell_F^2 - m)$. 

Here, the transmission and reflection coefficients, $T$ and $R$, are determined by matching the both spinor components of the solutions in each of the three regions:
\begin{align}
\begin{cases}
\psi_{\rm I }^{\uparrow}(x =  -\tfrac{\ell}{2}) = \psi_{\rm II }^{\uparrow}(x =  -\tfrac{\ell}{2}) \\ 
\psi_{\rm I }^{\downarrow}(x =  -\tfrac{\ell}{2}) = \psi_{\rm II }^{\downarrow}(x =  -\tfrac{\ell}{2}) \\
\psi_{\rm II }^{\uparrow}(x =  +\tfrac{\ell}{2}) = \psi_{\rm III }^{\uparrow}(x =  +\tfrac{\ell}{2}) \\ 
\psi_{\rm II }^{\downarrow}(x =  +\tfrac{\ell}{2}) = \psi_{\rm III }^{\downarrow}(x = +\tfrac{\ell}{2})
\end{cases}~. \label{eqFMatchCond}
\end{align}

To obtain the exact pair production rates, we need the explicit functions. Again, in the free regions I and III on either side of the capacitor, the eigenspinors correspond to plane-waves. Thus in the chiral basis we have,
\begin{align}
\psi_{\rm I}(x) &= e^{-i E t} \left[ 
\frac{e^{-i |k_-| x}}{\sqrt{2 |E_{-}| \ell_F}} \left( \begin{matrix} \sqrt{|E_-| + |k_-|} \\ \sqrt{|E_-| - |k_-|} \end{matrix} \right) \right. \nonumber\\
&\qquad \quad \left.+ R \frac{e^{+i |k_-| x}}{\sqrt{2 |E_{-}| \ell_F}} \left( \begin{matrix} \sqrt{|E_-| - |k_-|} \\ \sqrt{|E_-| + |k_-|} \end{matrix} \right)\right] ~, \label{eqFIR}\\
\psi_{\rm III}(x) &= e^{-i E t} \left[ 
T \frac{e^{i |k_+| x}}{\sqrt{2 E_{+} \ell_F}} \left( \begin{matrix} \sqrt{E_+ + |k_+|} \\ \sqrt{E_+ - |k_+|} \end{matrix} \right) \right] ~. \label{eqFT}
\end{align}
Note that the incident negative energy state in region-I and the transmitted positive-energy scattering state in region-III have the same chirality, while the reflected states have opposite chirality. From this, and the above expressions, it is clear that in the chiral limit the components of the reflected spinor will have smaller and smaller overlap with the incident spinor. Thus transmission should unitarize when $m \to 0$. This happens almost independently of the algebraic structure inherited from the harmonic oscillator differential equation.

However, it is instructive to see how the creation and annihilation structure of the differential equation in the field region acts to further trivialize and unitarize the transmission coefficient in the chiral limit. To do so, we give the explicit form of the spinor solutions to the Dirac equation~\eqref{eqODE1} in the intermediate region. They are given by
\begin{align}
\!\!\!\! \!\!\!\! 
&\psi_{\rm II}(x,t) = e^{-i E t} \left[ \sum_{\lambda = \pm} A_{\lambda} \psi_{\lambda}(x|E,\ell_F,m) \right] ~ , \\~  
&\psi_{\lambda}(x|E,\ell_F,m) := \left( \begin{matrix}
\qquad \quad ~-				f_{\lambda}(\chi,\mu) \\
\! \tfrac{1}{\mu} ( \chi + i \d_{\chi} )	f_{\lambda}(\chi,\mu)
\end{matrix} \right) ,
\!\!\!\! 
\label{eqFmid1}  \\
\!\!\!\! \!\!\!\! 
&f_{\pm}(\chi,\mu) := 
\sqrt{2}^{-\tfrac{1}{2} \mp \big(\tfrac{1}{2} + \tfrac{i}{2} \mu^2\big) } 
{\rm Exp}\left[- \tfrac{1}{2} (e^{\pm i \pi/4}\chi)^2\right]~\nonumber\\
&\qquad \qquad \qquad \qquad {\rm H}\!\left[-\tfrac{1}{2} \mp \big(\tfrac{1}{2} + \tfrac{i}{2} \mu^2\big) , e^{\pm i \pi/4} \chi \right],
\label{eqFermionSol}
\end{align}
where again $\mu = m \ell_F$, $\chi = (E \ell_F + x/\ell_F)$ and $\d_{\chi} = \ell_F \d_x$.

Further, replacing $\mu^2 \to 2n$ and $e^{i \pi/4} \chi \to x$ changes $f_{+}(\chi,\mu)$ into $\< x|n\> = {\rm H} [n,x]$, the $n^{th}$ energy eigenstate of the harmonic oscillator. Thus, in the chiral limit we see that $f_+(\chi,\mu) \to f_+(\chi,0)$ is simply the ground state for the harmonic oscillator. Now, the derivative operator in lower component is proportional to the annihilation operator of the harmonic oscillator. In the chiral limit, the oscillator structure of this integrable potential thus annihilates the lower component of $f_+(\chi,\mu)$.\footnote{One may worry about a $0/0$ ambiguity coming from the factor of $1/\mu$ in the lower component. This does not happen. Note that the eigenvalue of $(\chi + i \d_{\chi}) f_{+}(\chi,\mu)$ is proportional to $\mu^2$. So in the limit where $\mu \to 0$, the differential operator $\tfrac{1}{\mu} (\chi + i \d_{\chi}) f_{+}(\chi,\mu) \sim {\cal O}(\mu)$ and thus vanishes linearly as $m \to 0$.} Here, $\psi_{\pm}(x|E,\ell_F,m)$ is the wave-function for fermions with helicity $\pm \tfrac{1}{2}$ in the field-region. Here, in the chiral limit the lower component of $\psi_{+}(x)$ vanishes. 

Independent of the precise forms of the solutions in the field region, we can solve the boundary conditions in Eq.~\eqref{eqFMatchCond} by using the free-field solutions in Eqs.~\eqref{eqFIR} and~\eqref{eqFT}, and the schematic structure of the intermediate solution in Eq.~\eqref{eqFmid1}. Inputting these conditions yields the following transmission coefficient,

\begin{align}
T(E)_{\rm fermion} = \frac{4 i k_{-} e^{\tfrac{i}{2} \ell (k_{\rm I} - k_{\rm III})} \sqrt{\tfrac{|E_{+}|}{|E_{+}|} \tfrac{|E_{+}|-|k_{+}|}{|E_{-}|-|k_{-}|}}}{ {\cal D}f_{+}(+\tfrac{\ell}{2}) {\cal D}f_{-}(-\tfrac{\ell}{2}) - {\cal D}f_{-}(+\tfrac{\ell}{2}) {\cal D}f_{+}(-\tfrac{\ell}{2})} ~, \label{eqFermionT}
\end{align}
where ${\cal D}f(y)$ is the covariant-type derivative defined in Eq.~\eqref{eqDerivative}, and one of the factors of $2$ in the numerator comes from the Wronskian of $f_+$ and $f_-$. 

In the next section, we compare these exact tunneling rates for charged fermions and charged scalars to the Schwinger tunneling rate. We focus on strong electric fields, $m \ll 1/\ell_F$, whose length-scales are much larger than the width of the capacitor, $\ell \ll \ell_F$. In this regime, the potential varies extremely rapidly over the Compton wavelength of a virtual pair and the results from naive semiclassical analysis do not necessarily apply.

\subsection{Divergent scalar production at short distance}\label{secResult}

Tunneling and pair production rates in (1+1)-dimensions, $\tfrac{d^2 N}{dE dt}$, are related by~\cite{21-WangWong, 16-SchwingerRevisit},
\begin{align}
\frac{d^2 N}{dE dt} = \frac{v_R}{v_L} |T(E)|^2~,
\end{align}
where $v_R$ and $v_L$ are the phase velocities on the left and right of the field-region. The left-right asymmetry of the $v_L/v_R$ factor, here, corrects the inherent left-right asymmetry of the asymptotic scattering states in regions I and III. 
In the massless limit, this factor is identically one. In this section, we show that the exact tunneling rates for scalars in Eqs.~\eqref{eqScalarSol} and~\eqref{eqScalarT} diverge faster than any power-law in  $\ell_F/\ell$ in the limit where $m = 0$ and $\ell/\ell_F \to 0$. In contrast to this, the total pair production rate for massless fermions derived from Eqs.~\eqref{eqFermionSol} and~\eqref{eqFermionT} smoothly vanishes with $\ell$ when $\ell/\ell_F \to 0$.

\begin{figure}
\centering
\includegraphics[width=3.45in]{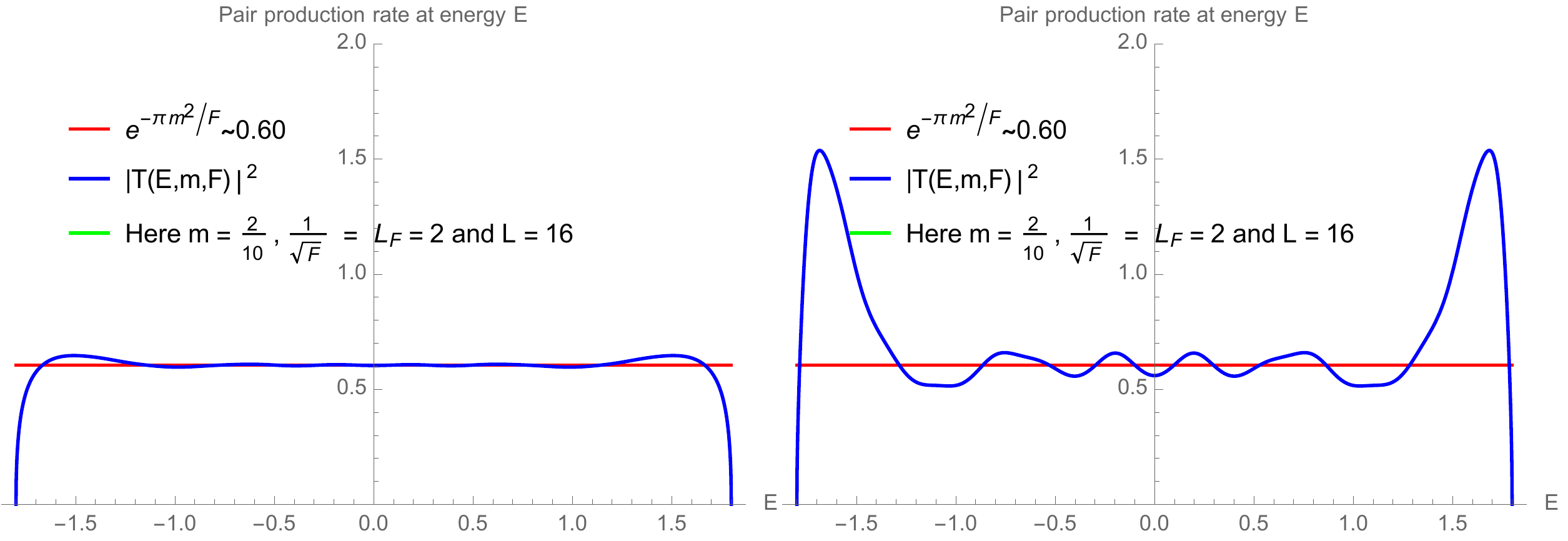}
\caption{Left: a plot of fermionic tunneling rates for states of energy $E$. Right: a plot of scalar tunneling rates for states of energy $E$. In these plots, we have $m = 2/10$, $\ell = 16$, and $\ell_F = 2$. Thus, $m \ell = 32/10 \sim 3 \gtrsim 1$ and $m \ell_F \sim 4/10 \lesssim 1$, while $\ell/\ell_F = 8 > 1$. For $|E| \simeq 0$, semiclassical analysis applies; here both results approximate the Schwinger result, with surpluses and deficits due to finite-distance corrections. Scalar pair production more than doubles near the edges, $|E| \sim \ell/2\ell_F^2-m$.}
\label{fig1}
\end{figure}

To make our case, we first study the exact rates for pair production per unit volume of phase-space. That is, we simply plot the rate of pair creation for very light charged scalars for various values of $\ell/\ell_F$, and for $m \ell_F \lesssim 1$. For comparison, we also show the fermionic pair production rates for the same values of the physical parameters. 

To begin, in Fig.~\ref{fig1} we plot the tunneling pair production for fermions and scalars for $m = 2/10, \ell = 16$ and $\ell_F = 2$. Here, the Schwinger pair creation rate is large but well less than unity: ${\rm Exp}[-\pi m^2/F] = {\rm Exp}[-\pi (m\ell_F)^2] \sim 0.60$. We can see that in the middle  of the pair creation regime, the fermionic and bosonic pair production rates approximately match, and oscillate around the Schwinger result. This is expected: For states in the middle of this band, the variation of the potential is slow and WKB/semiclassical analysis should give a good approximation to the exact physical tunneling rate. Further, we see nontrivial oscillations about the Schwinger result, coming from exact finite-distance corrections. These corrections become large near the edges of the mixing band. As observed in~\cite{21-WangWong}, the corrections for scalars more than double the Schwinger rate in these boundary regions.

\begin{figure}
\centering
\includegraphics[width=3.45in]{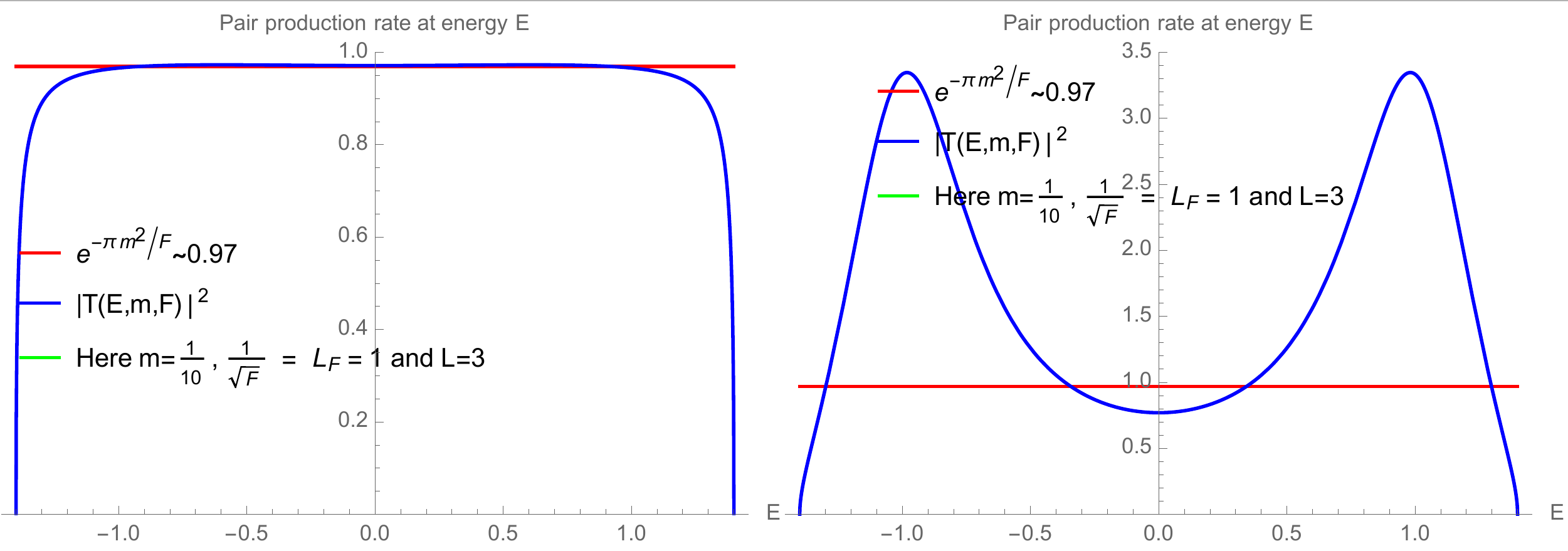}
\caption{Left: a plot of fermionic tunneling rates for states of energy $E$. Right: a plot of scalar tunneling rates for states of energy $E$. In these plots, we have $m = 1/10$, $\ell = 3$, and $\ell_F = 1$. Thus, $m \ell = 3/10 \lesssim 1$ and $m \ell_F = 1/10< 1 $, while $\ell/\ell_F = 3 \gtrsim 1$. Fermionic tunneling rates do not significantly exceed the Schwinger rate, while the bosonic overshoot it by a factor of $\sim3$ near the edges of the mixing window. Significant deviations begin to appear for scalars.}
\label{fig2}
\end{figure}

\begin{figure}
\centering
\includegraphics[width=3.45in]{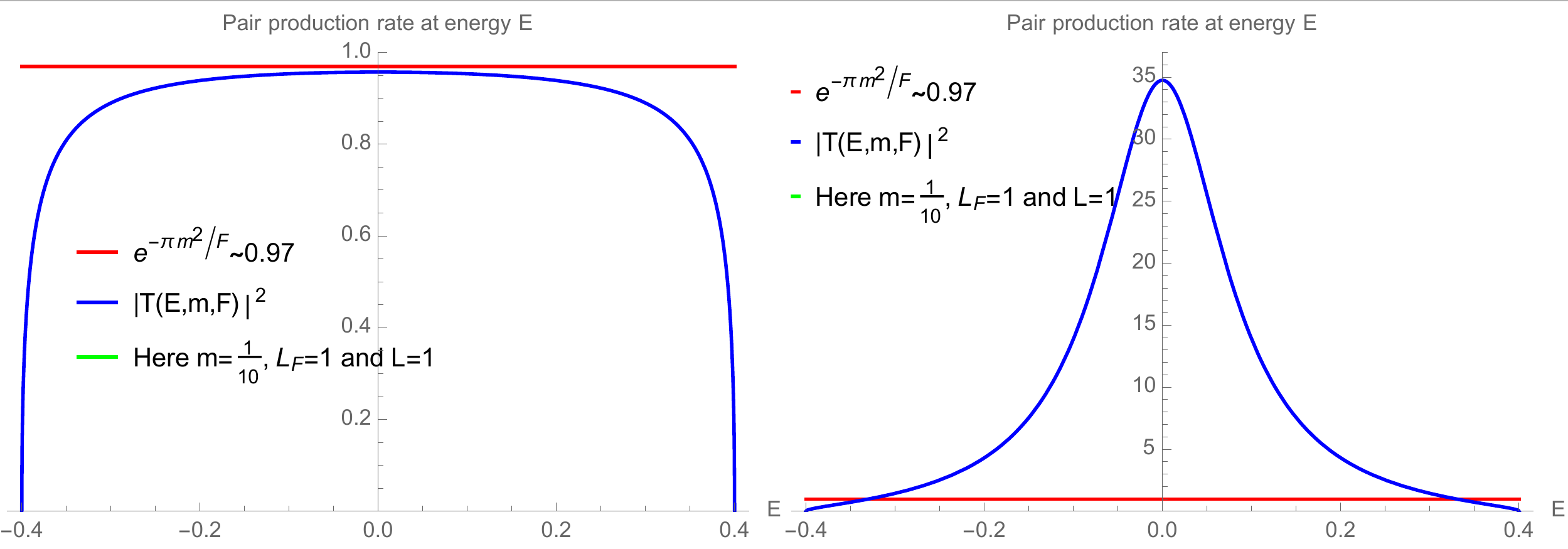}
\caption{Left: a plot of fermionic tunneling rates for states of energy $E$. Right: a plot of scalar tunneling rates for states of energy $E$. In these plots, we have $m = 1/10$, $\ell = 1$, and $\ell_F = 1$. Thus, $m \ell = 1/10 < 1$ and $m \ell_F = 1/10 < 1 $, while $\ell/\ell_F = 1$. Fermionic tunneling rates do not significantly exceed the Schwinger rate and nearly unitarize, while the bosonic overshoot it throughout the window, peaking at $\sim 35$ in the center.}
\label{fig3}
\end{figure}

\begin{figure}
\centering
\includegraphics[width=3.45in]{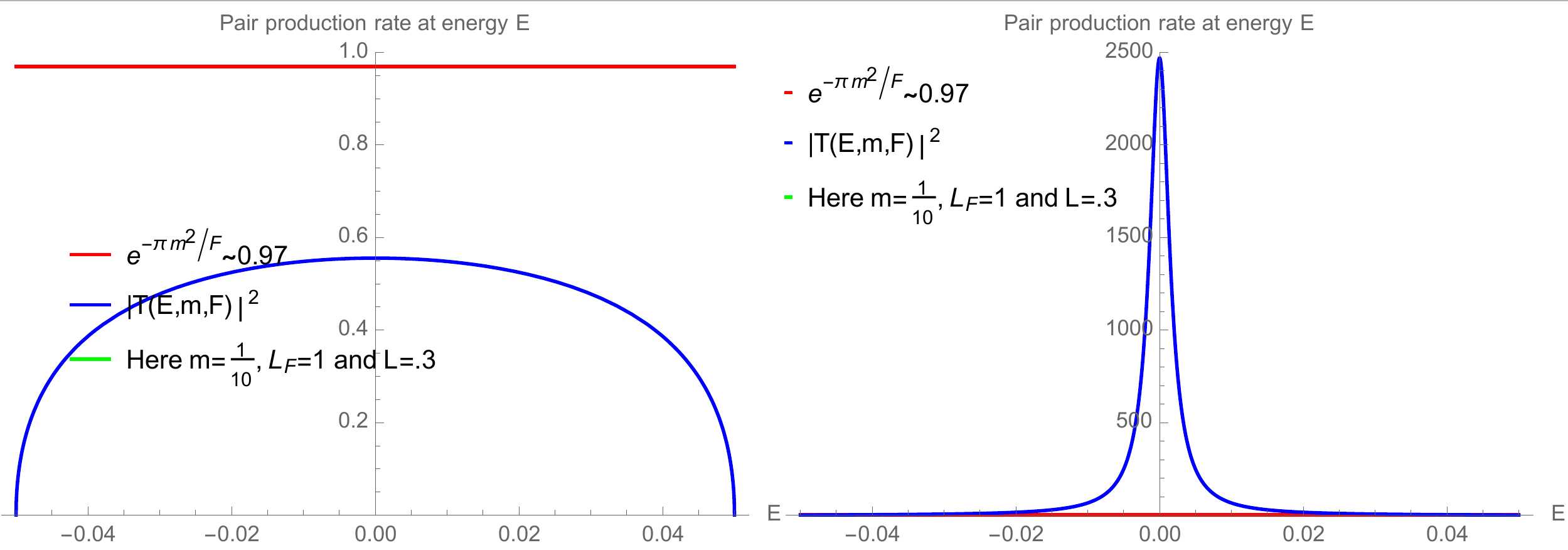}
\caption{Left: a plot of fermionic tunneling rates for states of energy $E$. Right: a plot of scalar tunneling rates for states of energy $E$. In these plots, we have $m = 1/10$, $\ell = 3/10$, and $\ell_F = 1$. Thus, $m \ell = 3/100 \ll 1$ and $m \ell_F = 1/10 < 1 $, while $\ell/\ell_F = 3/10 \lesssim 1$. Fermionic tunneling rates are significantly below the Schwinger rates due to velocity-blocking and Pauli-exclusion, while scalar production rates overshoot it throughout the mixing-window. At the center, scalar pair production rates exceed the Schwinger rate by a factor of $\sim2500$.}
\label{fig4}
\end{figure}

Continuing on, in Fig.~\ref{fig2}, Fig.~\ref{fig3}, and Fig.~\ref{fig4} we go closer to the massless limit, and study the charged fermion and charged scalar production for various widths of the field-region, $\ell$, while holding $m$ and $\ell_F$ fixed at $m = 1/10$ and $\ell_F = 1$. In Fig.~\ref{fig2}, we have $\ell = 3$ and see that the fermion tunneling amplitude nearly saturates the Schwinger result for the entire region of mixing. However, almost the entire mixing regime of the scalars is dominated by the excess ``shoulder region'' which more than triples the semiclassical Schwinger result. 

In Fig.~\ref{fig3}, we have $\ell = 1$ and thus $\ell/\ell_F = 1$. Here, the mixing region $m-\ell/2\ell_F^2 \leq E \leq \ell/2\ell_F^2 - m$ is entirely governed by the shoulder-region. For the fermions, the transmission coefficient is therefore sensitive to the fact that pair production ceases when $\ell/2\ell_F^2 \lesssim m$. Now, by helicity conservation in the chiral limit and Fermi-Dirac statistics that limit occupation numbers to be at most unity, the tunneling amplitude cannot exceed the Schwinger result. Finite-distance corrections for fermions therefore force the pair creation rates to be everywhere less than the Schwinger result. However, as discussed above, scalars are bound neither by helicity conservation nor Fermi-Dirac statistics to have occupation numbers less than unity. The shoulder-regions for scalar production in Fig.~\ref{fig3} can and do far exceed the Schwinger result, with maximum tunneling rates occurring in the middle-region where the combined effect of the two shoulder-excesses is greatest. This tunneling amplitude peaks at $\sim 35$ times the Schwinger result.

In Fig.~\ref{fig4}, we have $(\ell,\ell_F,m) = (3/10,1,1/10)$ and thus $\ell/\ell_F = 3/10 < 1$. The narrowness of the mixing region, $2 (\ell/2\ell_F^2 - m) = 1/10$ means that pair production is even further from the semiclassical result. For the fermions, the transmission coefficient is again everywhere less than the Schwinger result, as a consequence of Fermi-Dirac statistics and helicity conservation in the chiral limit. Surpluses above the Schwinger result in the shoulder-regions for scalars at $\ell/\ell_F = 3/10$ are even larger than at $\ell/\ell_F = 1$. Peak production rates occur in the middle of the shoulder region. They are $\sim 2500$ times the Schwinger rate for $\ell/\ell_F = 0.3$

\begin{figure}
\centering
\includegraphics[width=3.45in]{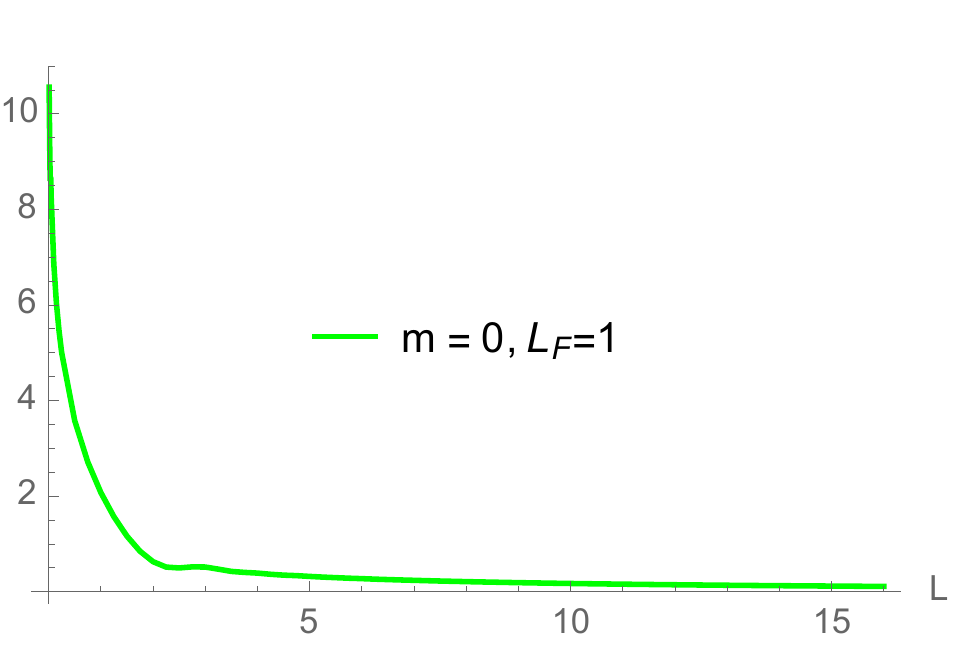}
\caption{A plot of the logarithm of the average pair production per unit length in the electric field region, $\log \overline{\Gamma}(m,\ell,\ell_F)$, for $m = 0$ and $\ell_F = 1$. We see that as $\ell/\ell_F \to 0$ this logarithm diverges to $+\infty$. Further, in the semiclassical limit $\ell \gg \ell_F$, we see that the logarithm approaches zero, and reproduces the Schwinger result: $\log \overline{\Gamma}(0,\ell,\ell_F) \to \log {\rm Exp}[-\pi (m\ell_F)^2] = -\pi (m\ell_F)^2 = 0$. The line in this graph connects numerically evaluated integrals at regular spacing; the minimal length computed is $\ell/\ell_F = 0.01$.}
\label{fig5}
\end{figure}

Finally we study the average pair-production per unit width in the mixed-energy regime $|E| \leq \ell/2 \ell_F^2 - m$, which we call $\overline{\Gamma}(m,\ell,\ell_F)$:
\begin{align}
\overline{\Gamma}(m,\ell,\ell_F) := \frac{1}{\ell/\ell_F^2 - 2m} \int_{m-\tfrac{\ell}{2\ell_F^2} }^{\tfrac{\ell}{2 \ell_F^2} - m} dE |T(E,m,\ell,\ell_F)|^2
\end{align} 
In Fig.~\ref{fig5} we show $\overline{\Gamma}(m,\ell,\ell_F)$ for scalars at $m = 0$ as a function of $x := \ell/\ell_F$, for $0.01 \leq x \leq 16$. To accurately show the growth rate, we are forced to make this a log-linear plot. From this, we surmise that scalar pair production clearly increases faster than any power-law in $\ell_F/\ell$ as $\ell/\ell_F \to 0$. Further, as expected we see that for $\ell/\ell_F \gtrsim 1$, we recover the Schwinger rate.

In summary, because interactions with electromagnetic fields preserve helicity, Fermi-Dirac statistics imply that charged fermion pair creation rates are bounded by the Schwinger rate in the chiral limit. Fermi-Dirac statistics, of course, do not apply to scalars. Further, scalars have neither helicity nor spinor components. Thus, no symmetry controls the rate of growth of the scalars in this limit. When we study the pair creation rate for massless charged scalars in a linear potential in the limit where the semiclassical approximation breaks-down, we find that the pair creation rate diverges.

We can understand this for the following reason. The standard description of the electric field generated by a pair of parallel capacitor plates is given by a piecewise continuous but not infinitely differentiable scalar potential. The cusps in its derivatives thus correspond to arbitrarily large energy-scales. These large energy scales may dominate any unprotected physical quantity, such as scalar pair creation rates, when all other energy-scales, chiefly the shift of the vacuum energy across the field-region, also become small. In the other direction, when the separation between the plates is large, the effects of these cusps are washed-out, and the semiclassical Schwinger result takes-over. In the next section, we discuss ways that this divergence is regulated in more realistic scenarios: smoothed potentials or massive scalars. 

\section{Regulating the divergence}\label{secSmoothing}

In this section, we discuss two natural physical mechanisms that regulate the divergence in charge-scalar pair production. First, in section~\ref{secMass} we show that if the scalars have masses, then the divergence is cut-off when $F \ell / m = \ell/m \ell_F^2 \sim 2$. However, before this point is reached, the rate of charged scalar pair creation grows faster than any power-law in $\ell_F/\ell$ for the regime $2\ell_F m \lesssim \ell/\ell_F \lesssim 3$. Second, in section~\ref{secTanh} we show evidence that smooth charge distributions do not lead to divergent rates of charged scalar pair production. In particular, we analyze the behavior of the exactly solvable tunneling amplitudes through the potential $w\ell_F^{-2} \tanh(x/w)$~\cite{18-Nikishov}. We analytically show that the tunneling rates, and therefore pair-production rates, are finite and regular in the limit where $w$ is sent to zero with $\ell_F$-fixed.

Each of these modifications to the standard description of a constant electric field between parallel capacitor plates smooths-out the jump-discontinuity at the wall of the conducting capacitor. This indicates that the divergence in the scalar production rates is strongly tied to the singularity in the second derivative of the vector potential, $A^{\mu}(x_{\alpha})$. Physically, gauge-invariant second-derivatives of $A^{\mu}$ correspond to current densities, $J_{\nu}(x_{\alpha}) = D^{\mu} F_{\mu \nu}(x_{\alpha})$. Thus, the model electric field from parallel-plate capacitors,
\begin{align}
&A_0(x)=   
\begin{cases} 
\ell/2\ell_F^2		& {\rm for} 	~ x \leq -\ell/2  \\ 
-x/\ell_F^2			& {\rm for} 	~ \leq x \leq \ell/2 \\ 
-\ell/2/\ell_F^2		& {\rm for} 	~ \ell/2 \leq x 
\end{cases} 
\nonumber\\
\implies 
&F(x)=   
\begin{cases} 
0 			& {\rm for} 	~ x\leq -\ell/2 \\ 
1/\ell_F^2 		& {\rm for} 	~ -\ell/2 \leq x \leq \ell/2 \\ 
0 			& {\rm for} 	~\ell/2 \leq x 
\end{cases} \label{eqCapPot1}~,
\end{align}
necessitate singular charge distributions $J_0(x) = \d_x F(x)$ given by, 
\begin{align}
J_0(x) = \ell/2\ell_F^2(\delta(x-\ell/2) - \delta(\ell/2-x))~.\label{eqCapDist1}
\end{align}
We present evidence tying the singularities of the $C^{0}$-potential in Eqs.~\eqref{eqCapPot1} and~\eqref{eqCapDist1} to divergent scalar production as $\ell/\ell_F \to 0$ in Figs.~\ref{fig5} and~\ref{fig6}.

\subsection{Masses}\label{secMass}

\begin{figure}
\centering
\includegraphics[width=3.45in]{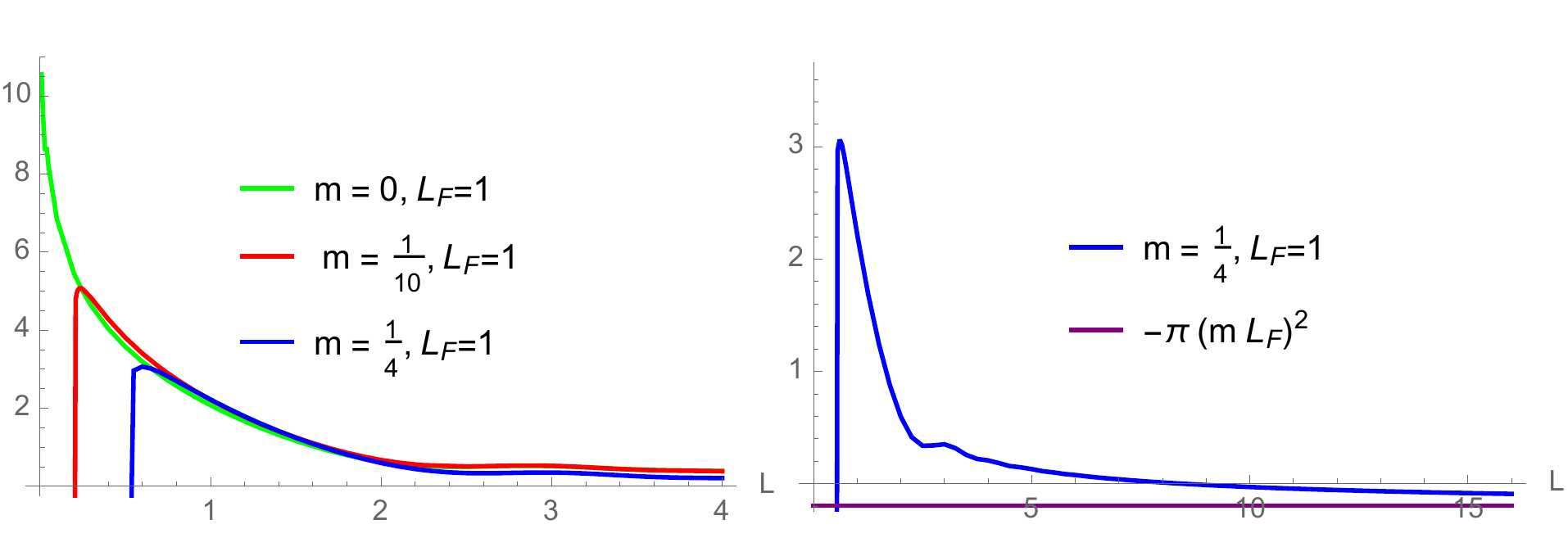}
\caption{Left: We plot of the logarithm of the average pair production per unit length in the field-region, $\log \overline{\Gamma}(m,\ell,\ell_F)$, for $\ell_F = 1$ and $m = 0, 1/10$ and $m = 1/4$. As $\ell/\ell_F \to 0$ the rate diverges exponentially quickly. Note that pair production rates are drastically different for $\ell \gg \ell_F$ where the tunneling is well approximated by the semiclassical result. However, when $\ell \lesssim 3 \ell_F$, we see that the pair production rates for scalars of each of the masses very closely follow the rate of massless pair production. Right: We show that $\log \overline{\Gamma}(m,\ell,\ell_F) = \log \overline{\Gamma}(1/4,\ell,1)$ smoothly approaches the Schwinger rate $\log {\rm Exp}[-\pi (m \ell_F)^2] = -\pi (1/4)^2 \sim -0.02$, for $\ell \gg \ell_F$.}
\label{fig6}
\end{figure}

Divergence at short distances are cut-off when the scalars have small but nonzero masses. In Fig.~\ref{fig6}, we plot the exact pair production rates for charged scalars for several fixed and small masses, as function of the dimensionless quantity $\ell/\ell_F$. Surprisingly, over the range $2m \ell_F^2 \lesssim \ell \lesssim 3 \ell_F$ \emph{massive scalar pair-production rates follow the pair-production rates for massless scalars}. When the separation drops to $\ell \sim 2m/\ell_F^2$, the constant electric field is incapable of forcing virtual vacuum fluctuations on-shell. Thus pair production ceases. Explicitly, as $\ell \to 2m \ell_F^2$ pair production ceases and $\log \overline{\Gamma}(m,\ell,\ell_F) \to -\infty$. 

Before closing, we briefly comment on three interesting features of the averaged pair-production rates in the range $2 m\ell_F^2 \lesssim \ell \lesssim 3\ell_F$. First, in Fig.~\ref{fig6} we see that the averaged pair production rates for both massive and massless scalars follow very similar trajectories when $\ell \lesssim 3 \ell_F$. This is entirely distinct to the behavior of the average pair production rates when $\ell \gtrsim 3 \ell_F$, where semiclassical tunneling rates govern pair creation and force the average pair production to be ${\rm Exp}[-\pi (m \ell_F)^2]$. At present, we do not understand why the tunneling amplitudes for scalars of different masses follow each other or why they follow the tunneling amplitudes for the massless scalars. 

Second, we note that this universal scaling can only happen in the strong field limit. This follows straightforwardly from the definition of this seemingly new scaling regime: $2 m\ell_F^2 \lesssim \ell \lesssim 3\ell_F$. For there to be any regime where the tunneling rates follow this new ``universal'' scaling, we must have $2 m \ell_F^2 \lesssim 3 \ell_F$. However, recalling $F = qE = \ell_F^{-2}$, we see that this immediately implies,
\begin{align}
2m \lesssim \frac{3}{\ell_F} \iff \left( \frac{2m}{3} \right)^2 \lesssim \frac{1}{\ell_F^2} = F~. \label{eqStrongUniverse}
\end{align}
Thus any study that focused on Schwinger pair creation in the weak-field regime would be insensitive to this seemingly new scaling.

Third, as we can see in both Figs.~\ref{fig5} and~\ref{fig6}, there is a small oscillation in $\log \overline{\Gamma}(m,\ell,\ell_F)$ when $\ell \sim 3 \ell_F$. For $\ell \gtrsim 3 \ell_F$, the tunneling is governed by the semiclassical result. However we can that the entire profile of the  tunneling amplitude changes in this regime. Specifically, if we compare Fig.~\ref{fig2} which plots the exact tunneling rate at $\ell/\ell_F = 3$ to Fig.~\ref{fig3} which plots $|T|^2$ for $\ell/\ell_F = 1$, we see that when $\ell/\ell_F \gtrsim 3$ the excess above the Schwinger result is confined to the shoulders of the mixing regime $|E| \leq \ell/2\ell_F^2 - m$. When this happens, the tunneling rate in these narrow shoulder regions only approximately triples the semiclassical result. However, when $\ell/\ell_F \lesssim 1$, the shoulders coalesce and the maximum excesses above the semiclassical result begins to increase exponentially. (For further evidence, see Fig.~\ref{fig4}.) Nevertheless, it is clear from the exact tunneling amplitudes that nothing singular is going on in these regions; the numerically obtained areas under the smooth $|T|^2$ plots in Figs.~\ref{fig2},~\ref{fig3} and~\ref{fig4} are well under control. The oscillations seen in Figs.~\ref{fig5} and~\ref{fig6} comes from the fact that the maximum production region shifts from the edges of the mixing region for $\ell \gtrsim 3 \ell_F$, to the center of the mixing region when $\ell \lesssim 3 \ell_F$.

\subsection{Smoothing the potential}\label{secTanh}

Neither chiral symmetry nor Fermi-Dirac statistics protect scalar pair production rates from diverging arbitrarily far from the semiclassical Schwinger result when the width of the field region $\ell$ is much smaller than the length-scale of the field itself ($\ell_F$). In other words semiclassical analysis no longer applies when $\ell/\ell_F \lesssim 1$. However, this divergence associated with ideal capacitor plates is likely inaccessible experimentally. To have the scalar potential~\eqref{eqCapPot1}, the charge distribution has to be exactly localized on the boundary of the capacitor plates as shown in Eq.~\eqref{eqCapDist1}. This is unlikely to occur in laboratory settings: charges \emph{smoothly} distribute themselves within the conducting capacitor plates.

It is natural to ask whether smooth charge distributions result in finite pair production rates in this short-distance limit. We now show that pair creation rates for an integrable $C^{\infty}$-deformation of Eq.~\eqref{eqCapPot1} are indeed finite. We study tunneling for the $\tanh$-potential:
\begin{align}
A_0(x) = \frac{\ell}{2 \ell_F^2} \tanh \left( \frac{x}{\ell/2} \right) 
\implies 
F(x) = \frac{1}{\ell_F^2} \sech^2 \left( \frac{x}{\ell/2} \right) ~.
\label{eqCapPot2}
\end{align}
We are specifically interested in whether the tunneling amplitudes $T$ for states in the mixing region $|E| \leq \ell/2\ell_F^2 - m$ either diverge or develop singularities when $\ell_F$ is fixed and $\ell \to 0$. 

To study this limit, we use Nikishov's exact result for scalar tunneling~\cite{16-SchwingerRevisit, 18-Nikishov},
\begin{align}
|T_{\rm scalar}^{\tanh}|^2 &= \frac{w}{1-w} \\
w:&\!= \frac{\sinh(2\pi \mu_{+}) \sinh(2\pi \mu_{-})}{\sinh^2(\pi(\mu_{+}+\mu_{-}))+\cosh^2(2 \pi t\lambda)}~,
\end{align}
where $t\lambda$ and $\mu_{\pm}$ are determined by the mass $m$ and energy $E$ of the scattering states, and by the force $1/\ell_F^2$ width of the field region $\ell$:
\begin{align}
&t\lambda:= \frac{1}{2} \sqrt{\left( \frac{\ell}{2 \ell_F} \right)^4 - 1~} \quad {\rm and} \\ 
&\mu_{\pm} :=  \frac{1}{4}\sqrt{\left(E \ell \pm \frac{\ell^2}{2\ell_F^2}\right)^2 - (m\ell)^2~}~.
\end{align}
Using Nikishov's explicit result, we can study its behavior in the two limiting regimes: $\ell/\ell_F \to \infty$ where the semiclassical Schwinger result should hold, and $\ell/\ell_F \to 0$ where the short-distance details of the potential dominate.

It is straightforward to show that the limit $\ell \to \infty$, for fixed $m$ and $\ell_F$, yields
\begin{align}
\lim_{\ell \to \infty} |T_{\rm scalar}^{\tanh}|^2 = e^{-\pi (m\ell_F)^2}~,
\end{align}
and exactly reproduces Schwinger's classic result~\cite{18-Nikishov}. 

Similarly, it is straightforward to show that the meromorphic function $|T_{\rm scalar}^{\tanh}|^2$ does not have any singularities on the real-$E$ axis in the limit where $\ell \to 0$ while $m$ and $\ell_F$ are held fixed. Notably, this conclusion holds even in the strict $m = 0$ limit. Explicitly, at $m = 0$, when we hold $\ell_F$-fixed, we have,
\begin{align}
\lim_{\ell \to 0} |T_{\rm scalar}^{\tanh}(m = 0)|^2 &= 
\lim_{\ell \to 0} \frac{(\pi E \ell)^2}{4} 
+ {\cal O}\left( \tfrac{E^2 \ell^4}{\ell_F^2},\tfrac{\ell^4}{\ell_F^4} \right) 
=0~.
\end{align}
Thus, we conclude that the $\tanh$-smoothed version of the naive capacitor-plate potential is free of divergences in the limit as the width of the field region shrinks to zero. 

The fact that charged-scalar pair production rates for the $\tanh$-potential in Eq.~\eqref{eqCapPot2} are finite indicates that the divergence for the capacitor-plate potential in Eq.~\eqref{eqCapPot1} is due to its singular charge distribution. Physically, while the idealization that charges distribute themselves on the outer-edge of a capacitor plate is a good approximation in practical laboratory settings, charges on physical plates have a finite but small penetration within the bulk near the boundary of a conducting metal plate. From this more practical perspective it is important to note that pair creation rates, even for massless scalars created by physical/smooth charge distributions, should be finite. 

It seems plausible that if the charge distribution penetrates only a finite depth $w$ into the material, say $\rho(\Delta x) \simeq {\rm Exp}[-\Delta x/w]$, then the scale $w$ should serve as an ultraviolet (UV) cutoff to the scalar tunneling amplitudes. If it serves as a cutoff in a similar manner to the masses in section~\ref{secMass}, then we may expect the scalar pair tunneling amplitudes to saturate at that IR scale,
\begin{align}
\bigg|T\bigg(\frac{\ell}{\ell_F}\bigg)\bigg|^2 \lesssim 
\bigg|T\bigg(\frac{ w }{\ell_F}\bigg)\bigg|^2 \sim 
{\rm Exp}\left[\frac{\ell_F}{w}\right]~,\!
\end{align}
It would be very interesting to test this in the future, for numerical (or exact) solutions to other smooth potentials.

Nonetheless, it is striking that light relativistic scalars are so incredibly sensitive to the shortest distance-scales in the problem. This is thrown into especially stark relief by the fact that light fermions are so strongly \emph{insensitive} to the short-distance structure of effective potentials. Fundamentally, it is extremely interesting (but perhaps not altogether surprising) that the distinction between these two phenomena comes directly from chiral symmetry and spin-statistics. In the final section, we comment on possible theoretical and experimental signatures of this short-distance behavior.

\section{Conclusions and future work}\label{secEnd}

Chiefly, we regard the contrast between the sensitivity of scalar production and the \emph{insensitivity} of fermionic pair production to the short-distance structure of the charge distribution $J_{\mu}(x_{\alpha})$ to be an amusing fact. It is quite natural that chiral symmetry should protect fermionic pair production from short-distance features. It is similarly natural that scalars are unprotected from violence in the ultraviolet, here represented by singular profiles of the current density $J_{\mu}(x_{\alpha})$. There are many directions for future work. We partition them first into building a better theoretical framework to understand these short-distance divergences, and second probing whether these short-distance effects can be visible in broader contexts.

\subsection{Theoretical framework}\label{secThy}

Schwinger pair creation can be mapped mode-by-mode to a scattering/tunneling problem in relativistic quantum mechanics in (1+1)-dimensions. In this sense, pair production is equivalent to a scattering phenomenon. Concrete computations however require one to make a gauge-choice. Recent advances in the S-matrix highlight the importance of understanding physical processes in manifestly gauge-invariant formulations~\cite{26-Smatrix1, 27-Smatrix2, 28-Smatrix3}. It would be very interesting to understand this precise S-matrix behavior in a manifestly gauge-invariant way. 

It is reasonable to suspect that manifestly gauge-invariant calculations of Schwinger pair creation are written in terms of non-local gauge-invariant contour integrals of the potential, such as Wilson lines. Indeed, the specific formulation in~\cite{25-Polyakov2} does something closely related to this. Expressing pair creation in terms of manifestly physical quantities, free from unphysical gauge redundancy, should help to highlight the analytic origin of the divergence in Fig.~\ref{fig5}.

Strikingly, the \emph{massive} scalar production curves in Fig.~\ref{fig6} almost exactly follow the \emph{massless} scalar production curves for $2 m \ell_F^2 \leq \ell \lesssim 3\ell_F$. This is starkly different to the behavior when $\ell \gg \ell_F$, where the Schwinger rates ${\rm Exp}[-\pi (m \ell_F)^2]$ cause exponential deviation. It would be extremely interesting to understand whether this similarity between dissimilar systems at short distances is universal and, if so, why. Again, a more general gauge-invariant formulation should allow us to divorce our analysis from the precise short-distance structure of the potential to see the underlying origin of the \emph{common} divergence between massive and massless scalars interacting with idealized parallel-plate capacitors.

In a different direction, the analogy between electric fields for idealized parallel-plate capacitors and harmonic oscillators is striking. It would be interesting to understand whether it can be used to explain the short-distance structure in scalar and fermion tunneling rates.

\subsection{Broader contexts}\label{secBroad}

The Schwinger mechanism is a simple example of both particle creation from an external field, and of back-reaction against this field. A more prominent example of this is Hawking radiation and Hawking evaporation. It would be interesting to study if the starkly distinct behavior of fermion vs. scalar Schwinger pair-creation rates at short-distances has an analog in Hawking radiation from very light and small Schwarzchild black holes. Is there an overproduction of scalars compared to fermions when the mass of a Schwarzchild black hole goes to zero?

Similarly, there is an idea that the smallness of the cosmological constant can be explained by Schwinger-like pair creation back-reacting against a large vacuum energy in de Sitter space~\cite{24-Polyakov1}. Here, too, it is natural to ask whether there is an analogously divergent rate of scalar pair production at short distances or early times. 

In a very different direction, the motivation for this paper was to study massless pair production from constant electric fields in condensed matter contexts. While the focus of this paper is tunneling phenomena in (1+1)-dimensions, it would be very interesting to see whether the range $\ell/\ell_F \lesssim 3$ is actually relevant for the proposed test of the Schwinger mechanism in graphene~\cite{03-SchwingerGraphene}.

A related question would be whether the suppressed fermionic tunneling amplitudes across short-distance electric fields can be used to create devices that measure short displacements with high accuracy. Although we did not emphasize it in the main body of the paper, the fermionic tunneling amplitudes shown in Figs.~\ref{fig2},~\ref{fig3}, and~\ref{fig4} are also very sensitive to small separations between parallel capacitor plates that have been given a fixed charge.

As commented in footnote 2, analytically continuing $E \to i E := B$ maps solutions to relativistic wave equations in pure electric fields to solutions to the same wave equations in pure magnetic fields. In this context, it is natural to ask if any interesting physics occurs in scaling regime $2m \ell_F^{2} \lesssim \ell \lesssim 3 \ell_F$ in Eq.~\eqref{eqStrongUniverse} when $\ell_F = (qB)^{-1/2}$ is the length scale of a pure magnetic field, $B$. Because magnetic fields do not do work, there is no pair creation here. Yet it is possible that strong deviations from the semiclassical picture shown in Fig.~\ref{fig6} may persist and have interesting features for the more stable configuration of supercritical magnetic fields. 

Further, it would be extremely interesting if the divergence in Fig.~\ref{fig6} were experimentally accessible in real world condensed-matter systems. In particular, it is possible that spin-charge separation that happens in some condensed matter systems could yield light charged-scalar degrees of freedom that obey Bose-Einstein statistics. In such a situation, it may be possible to observe this exponential divergence.

Finally, it is important to emphasize that these results do not apply only to the narrow topic of Schwinger pair creation. They apply to any situation where there are avoided level crossings that happen arbitrarily quickly. For example, one can reframe the Schwinger pair creation effect in terms of Landau-Zener avoided level-crossing between two states~\cite{11-LandauZener1, 12-LandauZener2}, one empty and the other filled, as they rapidly approach each other. 

As emphasized in Ref.~\cite{16-SchwingerRevisit}, if one goes to the gauge $E_i =F_{0i} = +\d_t A_i(t)$, then the semiclassical/Schwinger result is exactly the Landau-Zener tunneling amplitude. However, in this formulation the relevant physical picture is \emph{not} a scattering experiment but an avoided level-crossing. As avoided level crossings and the Landau-Zener effect are common throughout the condensed matter literature, it may be possible that the analogous limit to $\ell/\ell_F \to 0$ in this gauge maps to interesting condensed matter contexts.

\acknowledgements 

We thank Thomas Cohen, Poul Damgaard, Andrew Jackson, Michele Della Morte, Gordon Semenoff, and Cheuk-Yin Wong for discussions. The work of D.A.M. is supported by the Niels Bohr International Academy and a Carlsberg Distinguished Postdoctoral Fellowship (grant number CF-0183).

\end{document}